\documentclass[12pt]{article}

\usepackage{a4wide}
\usepackage{amsmath}
\usepackage{amssymb}
\usepackage{epsfig}

\newcommand{\s}{\hat{s}}

\newcommand{\Order}{{\mathcal O}}
\newcommand{\bsll}{b\to s\,\ell^{+}\ell^{-}}
\newcommand{\BXsll}{B\to X_s\,\ell^{+}\ell^{-}}

\newcommand{\eff}{\text{eff}}
\newcommand{\modi}{\text{mod}}
\newcommand{\pole}{\text{pole}}
\newcommand{\qr}{q\!\cdot\!r}

\newcommand{\rsl}{r \hspace{-5pt} / }

\newcommand{\Li}{\text{Li}}
\newcommand{\Brems}{\text{Brems}}
\newcommand{\wtC}{\widetilde{C}}
\newcommand{\wtO}{\widetilde{O}}

\newcommand{\st}{\rule[-1.5mm]{0mm}{6mm}}

\def\Journal#1#2#3#4{{#1} {\bf #2}, #3 (#4)}

\def\NPB{{\em Nucl. Phys.} B}
\def\PLB{{\em Phys. Lett.}  B}

\def\PRD{{\em Phys. Rev.} D}

\def\IJMP{\em Int. J. Mod. Phys.}
\begin{document}
\begin{titlepage}

  \begin{flushright}
    BUTP--02/06    \\
    hep-ph/0204341 \\[3ex]
  \end{flushright}
  \vspace{2.5cm}

  \begin{center}
    \LARGE{
        \bf Complete gluon bremsstrahlung corrections \\
        to the process $\boldsymbol{\bsll}$
        \footnote{Work partially supported by Schweizerischer Nationalfonds and SCOPES program}
    }
    \\[5ex]
    \large{
        H.H. Asatryan$^a$, H.M. Asatrian$^a$, C. Greub$^b$ and M. Walker$^b$
    }
    \\ \vspace*{0.3cm}
    \footnotesize
    {\it
        $^a$ Yerevan Physics Institute, 2 Alikhanyan Br., 375036 Yerevan, Armenia \\
        $^b$ Institut f\"ur Theoretische Physik, Universit\"at Bern, CH--3012 Bern, Switzerland.
    }
  \end{center}
  \vspace*{0.5cm}
  \begin{center}
    ABSTRACT \\
    \vspace*{0.5cm}
    \parbox{13cm}{
        In a recent paper \cite{virtCorr}, we presented the
        calculation of the $\Order(\alpha_s)$ virtual
        corrections to $\bsll$ and of those bremsstrahlung terms which are
        needed to cancel the infrared divergences.
        In the present paper we work out the remaining $\Order(\alpha_s)$
        bremsstrahlung corrections to $\bsll$, which do not suffer
        from infrared and collinear singularities. These new
        contributions turn out to be small numerically. In addition, we also investigate the
        impact of the definition of $m_c$ on the numerical results.
      }
  \end{center}

  \vfill
\end{titlepage}

\thispagestyle{empty}

%
%
\section{Introduction}
\label{sec:intro}The inclusive rare decay $\BXsll$ has not been observed so far, but is expected to be measured at the
operating $B$ factories after a few years of data taking. The measurement of its various kinematical distributions,
combined with improved data on $B\to X_s\,\gamma$, will imply tight constraints on the extensions of the standard model
and perhaps even reveal some new physics.

The main problem of the theoretical description of $\BXsll$ is due to the long-distance contributions from $\bar{c}c$
resonant states. When the invariant mass $\sqrt{s}$ of the lepton pair is close to the mass of a resonance, only model
dependent predictions for these long distance contributions are available today. It is therefore unclear whether the
theoretical uncertainty can be reduced to less than $\pm 20\%$ when integrating over these domains
\cite{Ligeti:1996yz}.

However, when restricting $\sqrt s$ to a region below the resonances, the long distance effects are under control. The
corrections to the pure perturbative picture can be analyzed within the heavy quark effective theory (HQET). In
particular, all available studies indicate that for the region $0.05<\hat{s}=s/m_b^2<0.25$ the non-perturbative effects
are below 10$\%$ \cite{Falk:1994dh,Ali:1997bm,chen:1997,Buchalla:1998ky,Buchalla:1998mt,Krueger:1996}. Consequently,
the differential decay rate for $\BXsll$ can be precisely predicted in this region using renormalization group improved
perturbation theory. It was pointed out in the literature that the differential decay rate and the forward-backward
asymmetry are particularly sensitive to new physics in this kinematical window
\cite{Ball,Lunghi,Silvestrini,ALGH,AAYS}.

The next-to-leading logarithmic (NLL) result for $\BXsll$ suffers from a relatively large ($\pm16\%$) dependence on the
matching scale $\mu_W$ \cite{Misiak:1993bc,Buras:1995dj}. The NNLL corrections to the Wilson coefficients remove the
matching scale dependence to a large extent \cite{Bobeth:2000}, but leave a $\pm13\%$-dependence on the renormalization
scale $\mu_b$, which is of $\Order(m_b)$. In order to further improve the theoretical prediction, we have recently
calculated the $\Order(\alpha_s)$ virtual two-loop corrections to the matrix elements $\langle s\, \ell^+ \ell^-|O_i|b
\rangle$ ($i=1,2$) as well as the virtual $\Order(\alpha_s)$ one-loop corrections to
$O_7$,...,~$O_{10}$~\cite{virtCorr}. As some of these corrections suffer from infrared and collinear singularities, we
have added those bremsstrahlung corrections needed to cancel these singularities. This improvement reduced the
renormalization scale dependence by a factor of 2.

In the present paper we complete the calculation of the bremsstrahlung corrections associated with the operators $O_1$,
$O_2$, $O_7$,...,$O_{10}$, i.e., we add those bremsstrahlung terms which are purely finite and have therefore been
omitted in Ref.~\cite{virtCorr}. We anticipate that the additional terms have a small impact on the phenomenology of
$\bsll$.

The paper is organized as follows: In Sec.~\ref{sec:effHamiltonian}, we briefly specify the theoretical framework,
before, in Sec.~\ref{sec:effcoeff}, we discuss the organization of the calculation of the finite bremsstrahlung
corrections and review the structure of the virtual corrections and singular bremsstrahlung contributions, calculated
in Ref.~\cite{virtCorr}. The finite bremsstrahlung corrections are worked out in Sec.~\ref{sec:O789} and
Sec.~\ref{sec:O1O2}. In Sec.~\ref{sec:numres}, finally, we investigate the numerical impact of the new corrections on
the invariant mass spectrum of the lepton pair. We also illustrate the dependence of our results on the definition of
the charm quark mass.
%
%
\section{Effective Hamiltonian}
    \label{sec:effHamiltonian}
The appropriate tool for studies on weak $B$ mesons decays is the effective Hamiltonian technique. The effective
Hamiltonian is derived from the standard model by integrating out the $t$ quark, the $Z_0$ and the $W$ boson. For the
decay channels $b\to s\, \ell^+\ell^-$ ($\ell=\mu,e$) it reads
\begin{equation*}
    \label{Heff}
    {\cal H}_{\eff} =  - \frac{4 \, G_F}{\sqrt{2}} \, V_{ts}^* \, V_{tb} \sum_{i=1}^{10} C_i \, O_i \, ,
\end{equation*}
where $O_i$ are dimension six operators and $C_i$ denote the corresponding Wilson coefficients. The operators we choose
as in \cite{Bobeth:2000}:
\begin{equation*}
    \label{oper}
    \begin{array}{rclrcl}
        O_1    & = & (\bar{s}_{L}\gamma_{\mu} T^a c_{L })
                     (\bar{c}_{L }\gamma^{\mu} T^a b_{L}), &
        O_2    & = & (\bar{s}_{L}\gamma_{\mu}  c_{L })
                     (\bar{c}_{L }\gamma^{\mu} b_{L}), \vspace{0.3cm}\\ \vspace{0.3cm}
        O_3    & = & (\bar{s}_{L}\gamma_{\mu}  b_{L })
                     \sum_q (\bar{q}\gamma^{\mu}  q), &
        O_4    & = & (\bar{s}_{L}\gamma_{\mu} T^a b_{L })
                     \sum_q (\bar{q}\gamma^{\mu} T^a q), \\ \vspace{0.2cm}
        O_5    & = & (\bar{s}_L \gamma_{\mu} \gamma_{\nu}
                     \gamma_{\sigma}b_L)
                     \sum_q(\bar{q} \gamma^{\mu} \gamma^{\nu}\gamma^{\sigma}q), &
        O_6    & = & (\bar{s}_L \gamma_{\mu} \gamma_{\nu}
                     \gamma_{\sigma} T^a b_L)
                     \sum_q(\bar{q} \gamma^{\mu} \gamma^{\nu}
                     \gamma^{\sigma} T^a q),    \vspace{0.2cm} \\ \vspace{0.2cm}
        O_7    & = & \frac{e}{g_s^2} m_b (\bar{s}_{L} \sigma^{\mu\nu}
                     b_{R}) F_{\mu\nu}, &
        O_8    & = & \frac{1}{g_s} m_b (\bar{s}_{L} \sigma^{\mu\nu}
                     T^a b_{R}) G_{\mu\nu}^a, \\ \vspace{0.2cm}
        O_9    & = & \frac{e^2}{g_s^2}(\bar{s}_L\gamma_{\mu} b_L)
                     \sum_\ell(\bar{\ell}\gamma^{\mu}\ell), &
        O_{10} & = & \frac{e^2}{g_s^2}(\bar{s}_L\gamma_{\mu} b_L)
                     \sum_\ell(\bar{\ell}\gamma^{\mu} \gamma_{5} \ell).
    \end{array}
\end{equation*}
The subscripts $L$ and $R$ refer to left- and right-handed fermion fields. We work in the approximation where the
combination $(V_{us}^* V_{ub})$ of Cabibbo-Kobayashi-Maskawa (CKM) matrix elements is neglected and the CKM structure
factorizes.

In the following it is convenient to define the operators $\widetilde{O}_7$,..., $\widetilde{O}_{10}$ according to
\begin{equation}
    \widetilde{O}_j = \frac{\alpha_s}{4\,\pi} \, O_j \, \, ,
    \quad (j=7,...,10) \, ,
\end{equation}
with the corresponding coefficients
\begin{equation}
    \widetilde{C}_j = \frac{4\,\pi}{\alpha_s} \, C_j \, \, , \quad (j=7,...,10) \, .
\end{equation}
%
%
\section{Organization of the calculation and previous results}
\label{sec:effcoeff}
\begin{figure}[t]
\begin{center}
    \includegraphics[width=8cm]{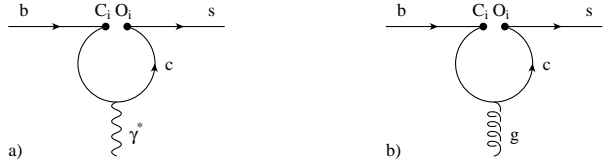}
    \caption{Diagram (a) can be absorbed by replacing the Wilson coefficients $\wtC_7$ and $\wtC_9$ through
    $\widetilde{C}_7^{\modi}$ and $\widetilde{C}_9^{\modi}$, respectively. $\gamma^*$ denotes an off-shell photon which
    subsequently decays into a $(\ell^+ \ell^-)$ pair. Similarly, diagram (b) is absorbed through the replacement
    $\widetilde{C}_8 \to \widetilde{C}_8^{\modi}$. $g$ denotes an on-shell gluon. The index $i$ runs from 1 to 6.
    See text for details.}
    \label{fig:diag1}
\end{center}
\end{figure}
\begin{figure}[t]
\begin{center}
    \includegraphics[width=\textwidth]{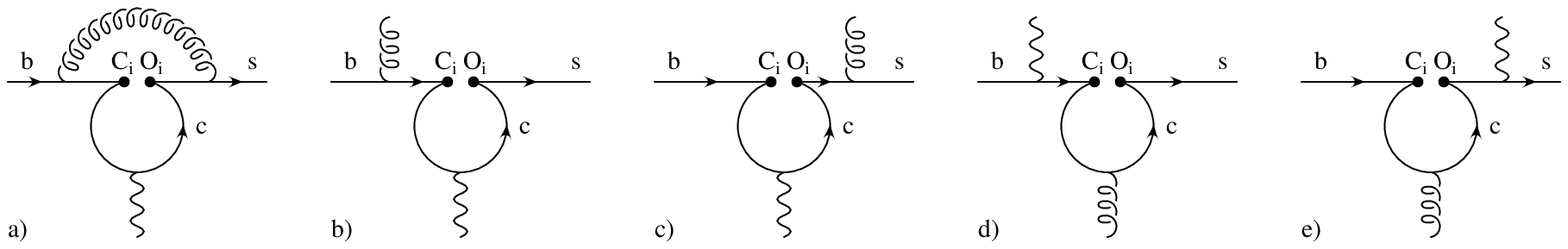}
    \caption{Diagrams which are automatically taken into account when calculating corrections to
        $\wtC_7^{(0,\modi)} \wtO_7$, $\wtC_8^{(0,\modi)} \wtO_8$ and $\wtC_9^{(0,\modi)} \wtO_9$.}
    \label{fig:diag2}
\end{center}
\end{figure}
\begin{figure}[t]
\begin{center}
    \includegraphics[width=\textwidth]{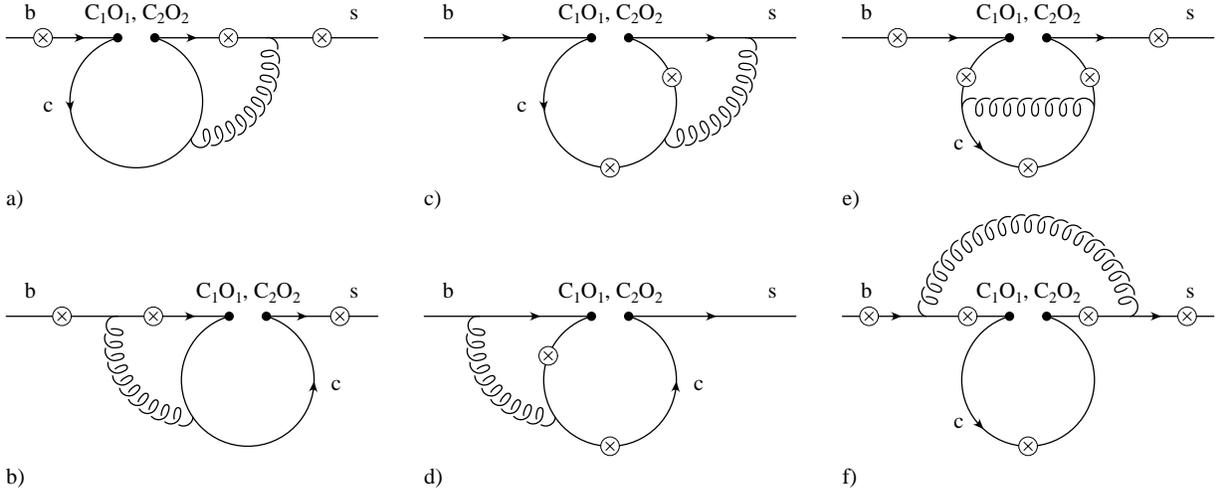}
    \caption{The two-loop virtual diagrams induced by $O_1$ and $O_2$ that cannot be absorbed into the
        $\widetilde{O}_{7,8,9}$ contributions by weighing them with the modified Wilson coefficients.
        The circle-crosses denote the possible locations where the virtual photon is emitted.
        The curly lines represent gluons.}
    \label{fig:diag3}
    \end{center}
\end{figure}
\begin{figure}[t]
\begin{center}
    \includegraphics[width=12cm]{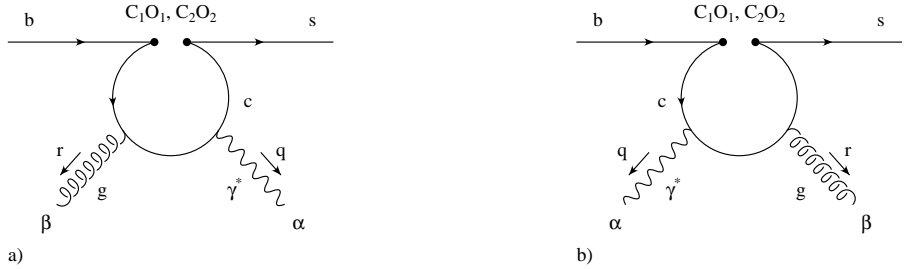}
    \caption{The only two bremsstrahlung diagrams induced by $O_1$ and $O_2$ that cannot be absorbed into the
        $\widetilde{O}_{7,8,9}$ contributions by weighing them with the modified Wilson coefficients.}
    \label{fig:diag4}
\end{center}
\end{figure}
\begin{figure}[t]
\begin{center}
    \includegraphics[width=\textwidth]{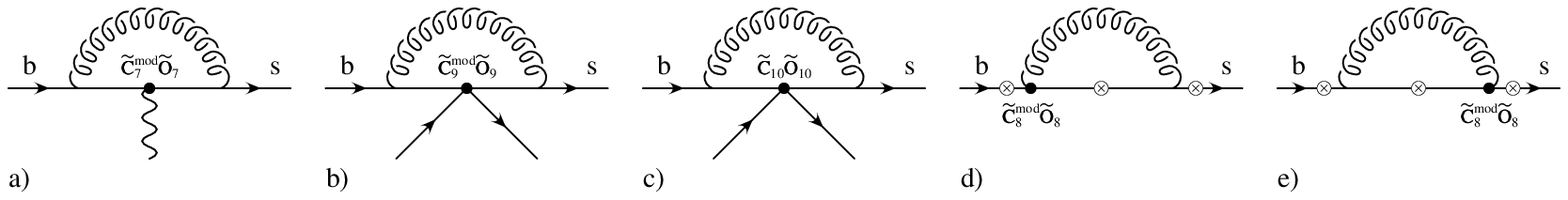}
    \caption{One-loop virtual $\Order(\alpha_s)$ corrections induced by
        $\widetilde{C}_7^{(0,\modi)} \widetilde{O}_7$,
        $\widetilde{C}_8^{(0,\modi)} \widetilde{O}_8$,
        $\widetilde{C}_9^{(0,\modi)} \widetilde{O}_9$
        and $\widetilde{C}_{10}^{(0)} \widetilde{O}_{10}$.
        The circle-crosses denote the possible locations for emission of a virtual photon.}
    \label{fig:diag5}
\end{center}
\end{figure}
\begin{figure}[t]
\begin{center}
    \includegraphics[width=\textwidth]{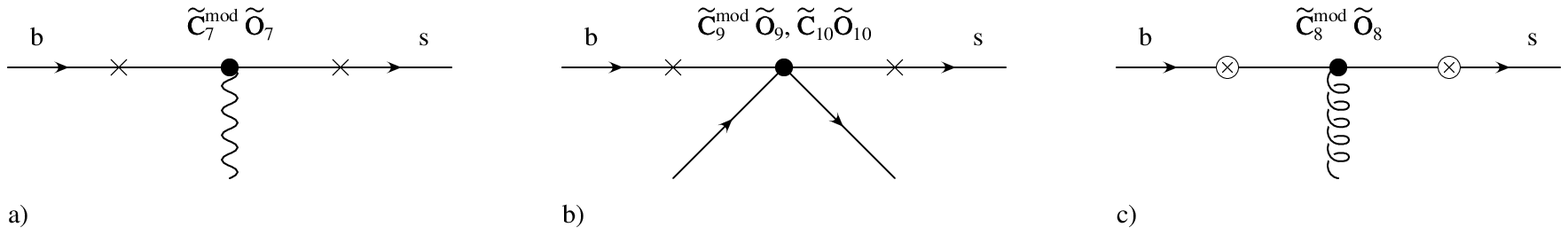}
    \caption{The $\Order(\alpha_s)$ bremsstrahlung diagrams induced by $\wtO_7$, $\wtO_9$, $\wtO_{10}$ and $\wtO_8$.
    Weighing the contributions of $\wtO_7$, $\wtO_8$ and $\wtO_9$ with the corresponding modified Wilson coefficients
    accounts for the bremsstrahlung diagrams depicted in Fig.~\ref{fig:diag2} (b)--(e).
    The crosses and circle-crosses denote the possible locations for emission of a bremsstrahlung gluon and a virtual photon, respectively.}
    \label{fig:diag6}
\end{center}
\end{figure}
In this section we comment on the organization of the calculation of the virtual and bremsstrahlung corrections to the
process $\bsll$ and repeat the results obtained in Ref.~\cite{virtCorr}.

The one-loop diagrams in Fig.~\ref{fig:diag1}, associated with the four-quark operators $O_1$,..., $O_6$, lead to
contributions which are proportional to the tree level matrix elements of the operators $\widetilde{O}_7$,
$\widetilde{O}_8$ and $\widetilde{O}_9$. Therefore, they can be absorbed by appropriately modifying the Wilson
coefficients $\widetilde{C}_7$, $\widetilde{C}_8$ and $\widetilde{C}_9$. The modified coefficients we write as
\begin{align}
        \label{modcoeff}
        \wtC_7^{\modi} = &\, A_7 , \\ \nonumber
        \wtC_8^{\modi} = &\, A_8 , \\
        \wtC_9^{\modi} = &\, A_9 + T_9 \, h(z,\s) + U_9 \, h(1,\s) + W_9 \, h(0,\s) \, . \nonumber
\end{align}
The auxiliary quantities $A_i$, $T_9$, $U_9$ and $W_9$ are linear combinations of the Wilson coefficients $C_i(\mu)$.
Their explicit form is relegated to the appendix. The one-loop function $h(z,\s)$ is given by \cite{Bobeth:2000}
\begin{multline}
\label{hfun}
    h(z,\s) = - \frac{4}{9} \ln(z) + \frac{8}{27} + \frac{16}{9}\frac{z}{\s} \\
              - \frac{2}{9} \left( 2+\frac{4\,z}{\s} \right)
              \sqrt{\left|\frac{4\,z-\s}{\s}\right|} \cdot
              \begin{cases}
                  2 \arctan \sqrt{\frac{\s}{4\,z-\s}} \,\, , & \s < 4\,z \\ \\
                  \ln \left(\frac{\sqrt{\s} + \sqrt{\s - 4\,z}}{\sqrt{\s} - \sqrt{\s - 4\,z}} \right) -i\,\pi, & \s > 4\,z
              \end{cases} \,\, .
\end{multline}
It is obvious that the modification of the Wilson coefficients automatically accounts also for the diagrams in
Fig.~\ref{fig:diag2} when calculating the corresponding corrections to the matrix elements
\[
    \langle s\, \ell^+ \ell^-|\widetilde{C}_i^{(0,\modi)} \widetilde{O}_i|b
    \rangle \quad (i=7,8,9),
\]
where $\widetilde{C}_i^{(0,\modi)}$ are the leading order terms of the modified Wilson coefficients, i.e.,
\begin{align}
\label{modcoeff0}
    \wtC_7^{(0,\modi)} = &\, A_7^{(0)} , \nonumber \\
    \wtC_8^{(0,\modi)} = &\, A_8^{(0)} , \\
    \wtC_9^{(0,\modi)} = &\, A_9^{(0)} + T_9^{(0)} \, h(z,\s) + U_9^{(0)} \, h(1,\s) + W_9^{(0)} \, h(0,\s) \, . \nonumber
\end{align}
For the explicit expressions of the quantities $A_i^{(0)}$, $T_9^{(0)}$, $U_9^{(0)}$ and $W_9^{(0)}$ we refer to the
appendix.

Notice that the virtual and bremsstrahlung corrections of the four-quark operators with topologies shown in
Figs.~\ref{fig:diag3} and \ref{fig:diag4}, however, have to be calculated explicitly. As the Wilson coefficients $C_1$
and $C_2$ are much larger than $C_3$,...,$C_6$ we retain the contributions of these topologies only for $O_1$ and $O_2$
insertions.

In the previous work \cite{virtCorr}, we systematically calculated the virtual corrections to the matrix elements of
$C_1^{(0)} O_1$, $C_2^{(0)} O_2$, shown in Fig.~\ref{fig:diag3}, as well as to those of $\widetilde{C}_j^{(0,\modi)}
\widetilde O_j$ ($j=7,...,9$) and $\widetilde{C}_{10}^{(0)} \widetilde O_{10}$ (cf Fig.~\ref{fig:diag5}). Furthermore,
we also took into account the corrections to the Wilson coefficients calculated in Refs.~\cite{Bobeth:2000,Chetyrkin}.

We found that the matrix elements of the operators $\widetilde{O}_7$, $\widetilde{O}_9$ and $\widetilde{O}_{10}$ [cf
Fig.~\ref{fig:diag5}(a)--(c)] suffer from infrared and collinear singularities. Consequently, on decay width level the
interferences $(\widetilde{O}_j, \widetilde{O}_k)$ ($j,k=7,9,10$) are singular, too. We therefore included the gluon
bremsstrahlung corrections associated with $(\widetilde{O}_j, \widetilde{O}_k)$ ($j,k=7,9,10$) in order to get an
infrared finite result for the decay width [cf Fig.~\ref{fig:diag6}(a) and (b)].

Taking into account the virtual and bremsstrahlung contributions discussed so far, we obtain the result presented in
Ref.~\cite{virtCorr}:
\begin{multline}
\label{rarewidth}
    \frac{d\Gamma(b\to X_s\, \ell^+\ell^-)}{d\s} =
    \left(\frac{\alpha_{em}}{4\,\pi}\right)^2
    \frac{G_F^2\, m_{b,\pole}^5\left|V_{ts}^*V_{tb}^{}\right|^2} {48\,\pi^3}(1-\s)^2 \\
    \times \left\{ \left (1+2\,\s\right) \left (\left |\widetilde C_9^{\eff}\right |^2 +
    \left |\widetilde C_{10}^{\eff}\right |^2 \right ) + 4\left(1+2/\hat s\right)\left |\widetilde C_7^{\eff}\right |^2 +
    12\, \mbox{Re}\left (\widetilde C_7^{\eff} \widetilde C_9^{\eff*}\right ) \right\},
\end{multline}
where the effective Wilson coefficients $\widetilde{C}_7^{\eff}$, $\widetilde{C}_9^{\eff}$ and
$\widetilde{C}_{10}^{\eff}$ are given by~\cite{virtCorr}

\begin{align}
    \label{effcoeff7}
    \widetilde C_7^{\eff} = & \left (1+\frac{\alpha_s(\mu)}{\pi} \,
        \omega_7 (\hat{s})\right ) \, A_7 \nonumber \\
        & -\frac{\alpha_{s}(\mu)}{4\,\pi}\left(C_1^{(0)} F_1^{(7)}(\s)+
        C_2^{(0)} F_2^{(7)}(\s) + A_8^{(0)} F_8^{(7)}(\s)\right),
    \\ \nonumber \\
    \label{effcoeff9}
    \widetilde C_9^{\eff} = & \left (1+\frac{\alpha_s(\mu)}{\pi} \,
        \omega_9 (\s)\right ) \left (A_9 + T_9 \, h (\hat m_c^2, \s)+U_9 \, h (1,\s) +
        W_9 \, h (0,\s)\right) \nonumber \\
        & -\frac{\alpha_{s}(\mu)}{4\,\pi}\left(C_1^{(0)} F_1^{(9)}(\s) + C_2^{(0)} F_2^{(9)}(\s)+
        A_8^{(0)} F_8^{(9)}(\s)\right),
    \\ \nonumber \\
    \label{effcoeff10}
    \widetilde C_{10}^{\eff} = & \left (1+\frac{\alpha_s(\mu)}{\pi} \,
        \omega_9 (\s)\right ) \, A_{10}.
\end{align}

The quantities $C_1^{(0)}$, $C_2^{(0)}$, $A_7$, $A_8^{(0)}$, $A_9$, $A_{10}$, $T_9$, $U_9$ and $W_9$ are Wilson
coefficients or linear combinations thereof. We give their analytical expressions and numerical values in the appendix.
The one-loop function $h (\hat m_c^2,\hat{s})$ is given in Eq.~(\ref{hfun}), while the two-loop functions
$F_{1,2}^{(7),(9)}$, accounting for the diagrams in Fig.~\ref{fig:diag3}, and the one-loop functions $F_8^{(7),(9)}$,
corresponding to the diagrams \ref{fig:diag5}(d) and (e), are given in Ref.~\cite{virtCorr}. The functions $\omega_7$
and $\omega_9$, finally, include both virtual and bremsstrahlung corrections associated with $\wtO_7$, $\wtO_9$ and
$\wtO_{10}$. For details on their construction we again refer to \cite{virtCorr}.

When calculating the decay width (\ref{rarewidth}), we retain only terms linear in $\alpha_s$ (and thus in  $\omega_7$,
$\omega_9$) in the expressions for $|\widetilde C_7^{\eff}|^2$, $|\widetilde C_9^{\eff}|^2$ and $|\widetilde
C_{10}^{\eff}|^2$. In the interference term $\text{Re} \left(\widetilde C_7^{\eff} \widetilde C_9^{\eff*}\right )$ too,
we keep only  linear contributions in $\alpha_s$. By construction one has to make the replacements $\omega_9 \to
\omega_{79}$ and $\omega_7 \to \omega_{79}$ in this term.

The functions $\omega_{7}$, $\omega_{9}$ and $\omega_{79}$ read
\begin{multline}
\omega_7(\s)  =
    - \frac{8}{3}\,\ln \left( \frac{\mu}{m_b} \right)
    - \frac{4}{3}\, \Li(\s)
    - \frac{2}{9} \,{\pi }^{2}
    - \frac{2}{3}\, \ln(\s) \ln(1-\s)
    \\
    - \frac{1}{3}\, \frac{8 + \s}{2 + \s} \ln (1-\s)
    - \frac{2}{3}\, \frac{\s \left(2 - 2\,\s - \s^2 \right)} {\left(1 - \s \right)^{2} \left(2 + \s \right)} \ln(\s)
    - \frac{1}{18}\, \frac {16 - 11\,\s - 17\,\s^{2}} {\left( 2 + \s \right) \left( 1 - \s \right)} \, , \\
\end{multline}
\begin{multline}
\omega_9(\hat{s}) =
    - \frac{4}{3}\, \Li(\s)
    - \frac{2}{3} \ln(1 - \s) \ln(\s)
    - \frac{2}{9}\pi^2
    - \frac{5 + 4\,\s}{3(1 + 2\,\s)} \ln(1-\s)
    \\
    - \frac{2\,\s\,(1 + \s)(1 - 2\,\s)}{3\,(1 - \s)^2 (1 + 2\,\s)} \ln(\s)
    + \frac{5 + 9\,\s - 6\,\s^2}{6\,(1 - \s)(1 + 2\,\s)} \, ,
\end{multline}
\begin{multline}
\omega_{79}(\s)  =
    - \frac{4}{3}\,\ln \left (\frac {\mu}{m_b}\right)
    - \frac{4}{3}\, \Li(\s)
    - \frac{2}{9}\,{\pi }^{2}
    - \frac{2}{3}\,\ln(\s) \ln(1 - \s)
    \\
    - \frac{1}{9}\,\frac{2 + 7\,\s}{\s} \ln (1-\s)
    - \frac{2}{9}\, \frac{\s \left(3 - 2\,\s \right)}{\left( 1 - \s \right)^{2}} \ln(\s)
    + \frac{1}{18}\, \frac{5 - 9\,\s}{1 - \s} \, .
\end{multline}

{\bf Summary} \\
The bremsstrahlung corrections associated with the interferences
\[
    \left( \wtC_j^{(0,\modi)} \wtO_j,\wtC_k^{(0,\modi)} \wtO_k \right), \quad (j,k=7,9,10),
\]
are already included in formula (\ref{rarewidth}). The remaining bremsstrahlung corrections, which are infrared finite,
we derive in Sec.~\ref{sec:O789} and Sec.~\ref{sec:O1O2}. In Sec.~\ref{sec:O789} we discuss the contributions of the
interferences
\[
    \left( \wtC_8^{(0,\modi)} \wtO_8,\wtC_k^{(0,\modi)} \wtO_k \right), \quad (k=7,8,9,10) \, ,
\]
which we call to be of type A. There is no contribution from $k=10$ because of the Dirac structures of the involved
operators. Sec.~\ref{sec:O1O2} is devoted to the interferences
\[
    \left( C_i^{(0)} O_i, C_j^{(0)} O_j \right), \,\, (i,j=1,2) \quad \text{and} \quad
    \left (C_i^{(0)} O_i, \wtC_k^{(0,\modi)} \wtO_k \right), \,\,
    (i=1,2;~k=7,8,9,10) \, .
\]
Accordingly, we call these the type B terms. Again, the contributions for $k=10$ vanish due to the Dirac structures of
the operators involved.
%
%
\section{Finite bremsstrahlung contributions of type A}
\label{sec:O789} The bremsstrahlung contributions taken into account by introducing the functions $\omega_i(\s)$ cancel
the infrared divergences associated with the virtual corrections. All other bremsstrahlung terms are finite. This
allows us to perform their calculation directly in $d=4$ dimensions.

The bremsstrahlung contributions from $\wtO_7-\wtO_8$ and
$\wtO_8-\wtO_9$
interference terms as well as the $\wtO_8-\wtO_8$ term oppose
no difficulties. The sum of these three parts can be written as
\begin{multline}
\label{finitea}
    \frac{d \Gamma^{\Brems,\text{A}}}{d\s}\, =
    \frac{d \Gamma_{78}^{\Brems}}{d\s}\, + \,\frac{d \Gamma_{89}^{\Brems}}{d\s}\, + \,\frac{d \Gamma_{88}^{\Brems}}{d\s}\,=\\
    \left(\frac{\alpha_{em}}{4\,\pi}\right)^2\, \left(\frac{\alpha_s}{4\,\pi} \right)
    \frac{m_{b,\pole}^5\,|V_{ts}^*\, V_{tb}|^2 \, G_F^2}{48\, \pi^3} \times
    \Big( 2\, \text{Re} \big[c_{78}\,\tau_{78} + c_{89}\,\tau_{89}\big] + c_{88}\,\tau_{88} \Big).
\end{multline}
The coefficients $c_{ij}$ are given by
\begin{equation}
    c_{78} = C_F \cdot \wtC_7^{(0,\eff)} \wtC_8^{(0,\eff)*}, \quad
    c_{89} = C_F \cdot \wtC_8^{(0,\eff)} \wtC_9^{(0,\eff)*} , \quad
    c_{88} = C_F \cdot \left| \wtC_8^{(0,\eff)} \right|^2 \, ,
\end{equation}
while the quantities $\tau_{ij}$ read
\begin{multline}
    \tau_{78} = \frac{8}{9\,\s} \,
    \Bigg\{\! 25 - 2\,\pi^2 - 27\,\s + 3\,\s^2 - \s^3 + 12 \left(\s+\s^2\right)\,\ln(\s) \\
    + 6 \left(\frac{\pi}{2}-\arctan\!\left[ \frac{2-4\,\s+\s^2}{(2-\s)\sqrt{\s}\,\sqrt{4-\s}} \right] \right)^2
    - 24 \, \text{Re} \! \left( \Li\!\left[\frac{\s-i\sqrt{\s}\,\sqrt{4-\s}}{2}\right] \right)- \\
    12\left((1-\s)\sqrt{\s}\,\sqrt{4-\s} - 2\arctan\!\left[ \frac{\sqrt{\s}\,\sqrt{4-\s}}{2-\s}\right] \right)
    \times \\
    \left( \arctan\!\left[ \sqrt{\frac{4-\s}{\s}}\,\right] -
        \arctan\!\left[ \frac{\sqrt{\s}\,\sqrt{4-\s}}{2-\s}\right] \right)
    \!\Bigg\} ,
\end{multline}

\begin{multline}
    \tau_{88} = \frac{4}{27\,\s} \,
    \Bigg\{\! -8\,\pi^2 + (1-\s)\left(77-\s-4\,\s^2\right)
    - 24\,\Li(1-\s) \\
    + 3 \left(10 - 4\,\s - 9\,\s^2 + 8\ln\!\left[\frac{\sqrt{\s}}{1-\s}\right]\right)\ln(\s)
    + 48\, \text{Re} \! \left( \Li\!\left[\frac{3-\s}{2} + i\, \frac{(1-\s)\sqrt{4-\s}}{2\,\sqrt{\s}} \right] \right) \\
    - 6 \left( \frac{20\,\s + 10\,\s^2 - 3\,\s^3}{\sqrt{\s}\,\sqrt{4-\s}} - 8\,\pi
    + 8 \arctan\!\left[ \sqrt{\frac{4-\s}{\s}}\,\right] \right) \times \\
    \left(\arctan\!\left[\sqrt{\frac{4-\s}{\s}}\,\right] -
        \arctan\!\left[\frac{\sqrt{\s}\,\sqrt{4-\s}}{2-\s}\right] \right)
    \! \Bigg\} ,
\end{multline}

\begin{multline}
    \tau_{89} = \frac{2}{3} \,
    \Bigg\{ \s\,(4-\s) - 3 - 4\,\ln(\s) \big(1-\s-\s^2\big) \\
    - 8\, \text{Re} \!
    \left( \Li\! \left[ \frac{\s}{2} + i\,\frac{\sqrt{\s}\,\sqrt{4-\s}}{2} \right]
    -  \Li\! \left[ \frac{-2 + \s(4-\s)}{2} + i \, \frac{(2-\s) \sqrt{\s}\,\sqrt{4-\s}}{2} \right] \right) \\
    + 4 \left( \s^2\, \sqrt{\frac{4-\s}{\s}} + 2\,\arctan\!\left[ \frac{\sqrt{\s}\,\sqrt{4-\s}}{2-\s}\right] \right)
    \times \\
    \left( \arctan\!\left[ \sqrt{\frac{4-\s}{\s}}\,\right] -
        \arctan\!\left[ \frac{\sqrt{\s}\,\sqrt{4-\s}}{2-\s}\right]\right)
    \Bigg\} .
\end{multline}
\newpage
%
%
\section{Finite bremsstrahlung contributions of type B}
\label{sec:O1O2} In this section we consider the bremsstrahlung  contributions from $O_1$ and $O_2$ and interference
terms with $\wtO_7$, $\wtO_8$, $\wtO_9$ and $\wtO_{10}$. As mentioned before, interferences with $\wtO_{10}$ vanish due
to the Dirac structures of the operators.

The bremsstrahlung contributions discussed in this section all involve the matrix elements associated with the two
diagrams depicted in Fig.~\ref{fig:diag4}. Their sum, $\bar{J}_{\alpha \beta}$, is given by
\begin{multline}
\label{bb2}
    \bar{J}_{\alpha\beta} = \frac{e\,g_s\,Q_u}{16\,\pi^2}
    \left[
        E(\alpha,\beta,r) \, \bar\Delta i_5 +
        E(\alpha,\beta,q) \, \bar\Delta i_6 -
        E(\beta,r,q)\frac{r_{\alpha}}{\qr} \, \bar\Delta i_{23}
    \right.\\
    \left.
        - E(\alpha,r,q)\frac{q_{\beta}}{\qr} \, \bar\Delta i_{26}
        - E(\beta,r,q)\frac{q_{\alpha}}{\qr} \, \bar\Delta i_{27}
    \right]
    L \, \frac{\lambda}{2} \, ,
\end{multline}
where $q$ and $r$ denote the momenta of the virtual photon and of the gluon, respectively. The index $\alpha$ will be
contracted with the photon propagator, whereas $\beta$ is contracted with the polarization vector $\epsilon^\beta(r)$
of the gluon. $\bar J_{\alpha\beta}$ and $\bar{\Delta}i_k$ are obtained from $J_{\alpha\beta}$ and $\Delta i_k$
\cite{virtCorr}, respectively, by setting $r^2=0$ and dropping terms proportional to $r_\beta$. The matrix
$E(\alpha,\beta,r)$ is defined as
\begin{equation}
    E(\alpha,\beta,r) = \frac{1}{2} (\gamma_{\alpha}\gamma_{\beta}\rsl - \rsl\gamma_{\beta}\gamma_{\alpha}).
\end{equation}
Due to Ward identities, the quantities $\bar\Delta i_k$ are not independent of one another. Namely,
\[
    q^\alpha \bar{J}_{\alpha\beta} = 0 \quad \text{and} \quad r^\beta \bar{J}_{\alpha\beta} = 0
\]
imply that $\bar\Delta i_5$ and $\bar\Delta i_6$ can be expressed as
\begin{equation}
    \bar\Delta i_5 = \bar\Delta i_{23} + \frac{q^2}{\qr} \bar\Delta i_{27} \, ; \quad
    \bar\Delta i_6 = \bar\Delta i_{26}\, .
\end{equation}
As in addition $\bar\Delta i_{26} = - \bar\Delta i_{23}$, the bremsstrahlung matrix elements depend on $\bar\Delta
i_{23}$ and $\bar\Delta i_{27}$, only. In $d=4$ dimensions we find
\begin{eqnarray}
    \label{eq:deltaik}
    \bar\Delta i_{23} &=& 8\, (\qr) \int_0^1\! dx\,dy\, \frac{x\, y (1-y)^2}{C} \, ,
    \nonumber \\
    \bar\Delta i_{27} &=& 8\, (\qr) \int_0^1\! dx\,dy\, \frac{y\, (1-y)^2}{C} \, ,
\end{eqnarray}
where
\begin{equation*}
    C = m_c^2 - 2\, x\, y (1-y) (\qr) - q^2\, y\, (1-y) - i\,\delta.
\end{equation*}

In the rest frame of the $b$ quark and for fixed $\s=q^2/m_b^2$, the phase space integrals which one encounters in the
calculation of $d \Gamma^{\Brems,\text{B}}/d\s$ can be reduced to a two-dimensional integral over $\hat{E}_r = E_r/m_b$
and $\hat{E}_s = E_s/m_b$, where $E_r$ and $E_s$ are the energy of the gluon and the $s$ quark, respectively. In the
following it is useful to introduce the integration variable $w = 1 - 2 \hat{E}_s$ instead of $\hat{E}_s$. The
integration limits are then given by
\begin{equation*}
    \hat{E}_r \in \left[ \frac{w-\s}{2},  \frac{w-\s}{2\,w}\right]
\quad \text{and} \quad w \in [\s,1].
\end{equation*}
For fixed values of $\s$, the quantities $\bar\Delta i_{23}$ and $\bar\Delta i_{27}$ depend only on the scalar product
$\qr$, which, in the rest frame of the $b$ quark, is given by $(w-\s)\,m_b^2/2$. The integration over $\hat{E}_r$ turns
out to be of rational kind and can be performed analytically. The remaining integration over $w$, however, is more
complicated and is done numerically. The result can be written as
\begin{multline}
\label{eq:tau}
    \frac{d \Gamma^{\Brems,\text{B}}}{d\s}\, =
    \left( \frac{\alpha_{\text{em}}}{4\,\pi} \right)^2 \left( \frac{\alpha_s}{4\,\pi} \right)
    \frac{G_F^2\, m_{b,\pole}^5 \,|V_{ts}^*\, V_{tb}|^2}{48\,\pi^3} \times \\
    \int\limits_{\s}^{1}\!d w\,
    \Big[
        \left( c_{11} + c_{12} + c_{22} \right) \tau_{22} +
        2\,\text{Re} \big[
        \left( c_{17} + c_{27} \right) \tau_{27} +
        \left( c_{18} + c_{28} \right) \tau_{28} +
        \left( c_{19} + c_{29} \right) \tau_{29}
        \big]
    \Big].
\end{multline}
The quantities $\tau_{ij}$, expressed in terms of
$\bar\Delta i_{23}$ and $\bar\Delta i_{27}$, read
\begin{multline}
    \tau_{22} = \frac{8}{27} \frac{(w-\s)(1-w)^2}{\s\,w^3}
    \times
    \bigg\{
        \Big[ 3\,w^2 + 2\,\s^2(2+w) - \s\,w\,(5-2\,w) \Big] \left| \bar\Delta i_{23} \right|^2 + \\
        \Big[ 2\,\s^2\,(2+\,w) + \s\,w\,(1+2\,w) \Big] \left| \bar\Delta i_{27} \right|^2 +
        4\,\s \Big[w\,(1-w) - \s\,(2+w) \Big] \cdot \text{Re}\left[ \bar\Delta i_{23} \bar\Delta i_{27}^* \right]
    \bigg\}
\end{multline}

\begin{multline}
    \tau_{27} = \frac{8}{3}\frac{1}{\s\,w} \times
    \bigg\{\Big[
        (1-w)\left(4\,\s^2 - \s\,w+w^2\right) +
            \s\,w\,(4+\s-w) \ln(w) \Big] \bar\Delta i_{23} \\
        - \Big[ 4\,\s^2\,(1-w) + \s\,w\,(4 + \s -w)\,\ln(w) \Big] \bar\Delta i_{27}
    \bigg\}
\end{multline}

\begin{multline}
    \tau_{28} = \frac{8}{9}\frac{1}{\s\,w\,(w-\s)} \times
    \Bigg\{\!
        \Big[(w-s)^2(2\,\s-w)(1-w) \Big] \bar\Delta i_{23}
        - \Big[ 2\,\s\,(w-\s)^2(1-w) \Big] \bar\Delta i_{27} \\
        + \s\,w \Big[ (1+2\,\s-2\,w)\bar\Delta i_{23} - 2\, (1+\s-w) \bar\Delta i_{27}\Big] \cdot
        \ln\!\left[ \frac{\s}{(1+\s-w)(w^2+\s\,(1-w))} \right]
    \Bigg\}
\end{multline}

\begin{multline}
    \tau_{29} = \frac{4}{3}\frac{1}{w}\times
    \bigg\{
        \Big[ 2\,\s(1-w)(\s+w) + 4\,\s\,w\ln(w) \Big] \bar\Delta i_{23} - \\
        \Big[ 2\,\s(1-w)(\s+w) + w (3\,\s+w) \ln(w) \Big] \bar\Delta i_{27}
    \bigg\}
\end{multline}
The coefficients $c_{ij}$ in Eq.~(\ref{eq:tau}) include the dependence on the Wilson coefficients and the color
factors. Explicitly, they read
\begin{align}
    c_{11} = & \, C_{\tau_1} \cdot \left|C_1^{(0)}\right|^2, &
    c_{17} = & \, C_{\tau_2} \cdot C_1^{(0)} \wtC_7^{(0,\eff)*}, &
    c_{27} = & \, C_F \cdot C_2^{(0)} \wtC_7^{(0,\eff)*}, \nonumber \\
    c_{12} = & \, C_{\tau_2} \cdot 2\,\text{Re}\!\left[ C_1^{(0)} C_2^{(0)*}\right], &
    c_{18} = & \, C_{\tau_2} \cdot C_1^{(0)} \wtC_8^{(0,\eff)*}, &
    c_{28} = & \, C_F \cdot C_2^{(0)} \wtC_8^{(0,\eff)*}, \\
    c_{22} = & \, C_F \cdot \left|C_2^{(0)}\right|^2, &
    c_{19} = & \, C_{\tau_2} \cdot C_1^{(0)} \wtC_9^{(0,\eff)*}, &
    c_{29} = & \, C_F \cdot C_2^{(0)} \wtC_9^{(0,\eff)*}, \nonumber
\end{align}
where the color factors $C_F$, $C_{\tau_1}$ and $C_{\tau_2}$ arise from the following color structures:
\[
    \sum_a T^a T^a = C_F \boldsymbol{1}, \quad C_F = \frac{N_c^2-1}{2\, N_c}\, ,
\]
\[
    \sum_{a,b,c} T^a T^c T^a T^b T^c T^b = C_{\tau_1} \boldsymbol{1}, \quad C_{\tau_1} = \frac{N_c^2-1}{8\,N_c^3} \, ,
\]
and
\[
    \sum_{a,b} T^a T^b T^a T^b  = C_{\tau_2} \boldsymbol{1}, \quad C_{\tau_2} = -\frac{N_c^2-1}{4\,N_c^2}\, .
\]
Finally, we list the explicit formulas for $\bar\Delta i_{23}$ and $\bar\Delta i_{27}$ expressed as a function of $\s$
and the integration variable $w$. We obtain
\begin{align}
    \bar\Delta i_{23} & = -2 + \frac{4}{w-\s}
    \left[ z\, G_{-1}\!\left(\frac{\s}{z}\right) - z\, G_{-1}\!\left(\frac{w}{z}\right)
    - \frac{\s}{2}\, G_0\!\left(\frac{\s}{z}\right) + \frac{\s}{2}\,G_0\!\left(\frac{w}{z} \right) \right], \\
    \nonumber \\
    \bar\Delta i_{27} & = 2 \left[ G_0\!\left(\frac{\s}{z}\right) - G_0\!\left(\frac{w}{z}\right)
    \right],
\end{align}
where the functions $G_k(t)$ ($k \ge -1$) are defined through the integral
\[
    G_k(t) = \int\limits_0^1\! d x \, x^k \, \ln\left[ 1-t\,x(1-x)-i\,\delta\right],
        \quad G_1(t) = \frac{1}{2} G_0(t).
\]
Explicitly, the functions $G_{-1}(t)$ and $G_0(t)$ read
\begin{align}
    G_{-1}(t) & =
    \begin{cases}
        2\,\pi\, \arctan\!\left( \sqrt\frac{4-t}{t}\,\right) - \frac{\pi^2}{2}
            - 2\,\arctan^2\!\left( \sqrt\frac{4-t}{t}\,\right),
            & t < 4 \\ \\
        - 2\,i\,\pi\,\ln\!\left( \frac{\sqrt{t}+\sqrt{t-4}}{2}\right) - \frac{\pi^2}{2}
            + 2\, \ln^2\!\left( \frac{\sqrt{t}+\sqrt{t-4}}{2}\right),
            & t > 4
    \end{cases}\quad, \\ \nonumber \\
    G_0(t) & =
    \begin{cases}
        \pi \, \sqrt{\frac{4-t}{t}} - 2 - 2\, \sqrt{\frac{4-t}{t}}\, \arctan\!\left( \sqrt\frac{4-t}{t}\,\right),
            & t < 4 \\ \\
        -i\,\pi \sqrt{\frac{t-4}{t}} - 2 + 2\, \sqrt{\frac{t-4}{t}}\,\ln\!\left( \frac{\sqrt{t}+\sqrt{t-4}}{2}\right),
            & t > 4
    \end{cases}\quad.
\end{align}
%
%
\section{Numerical results}
    \label{sec:numres}
First, we   investigate the numerical impact of the finite bremsstrahlung corrections [see Eqs.~(\ref{finitea} and
(\ref{eq:tau})] on the dilepton invariant mass spectrum. Following common practice, we consider the ratio
\begin{equation}
\label{rdiff}
    R_{\text{quark}}(\s) = \frac{1}{\Gamma(b\to X_c\,e\,\bar\nu_e)}
                           \frac{d\Gamma(\bsll)}{d\s} \, ,
\end{equation}
in which the factor $m_{b,\pole}^5$ drops out. The explicit expression for the semi-leptonic decay width $\Gamma(b \to
X_c\,e\,\bar\nu_e)$ reads
\begin{equation}
    \label{widthsl} \Gamma(b \to X_c\, e\,\bar{\nu}_e) = \frac{G_F^2 \, m_{b,\pole}^5}{192\, \pi^3} \, |V_{cb}|^2 \, g \!
    \left( \frac{m_{c,\pole}^2}{m_{b,\pole}^2} \right) \, K \! \left( \frac{m_c^2}{m_b^2} \right) \, ,
\end{equation}
where $g(z)=1-8 \,z +8 \, z^3 - z^4 -12 \, z^2 \, \ln(z) \, $
is the phase space factor, and
\begin{equation}
    K(z) = 1 - \frac{2\, \alpha_s(m_b)}{3\,\pi} \, \frac{f(z)}{g(z)}
\end{equation}
incorporates the next-to-leading QCD correction to the semi-leptonic decay \cite{Cabibbo}. The function $f(z)$ has been
calculated analytically in Ref.~\cite{Nir}. It reads
\begin{align}
\label{ffun}
    f(z) = & -(1-z^2) \, \left( \frac{25}{4} - \frac{239}{3} \, z + \frac{25}{4} \, z^2 \right) + z \, \ln(z)
        \left( 20 + 90 \, z -\frac{4}{3} \, z^2 + \frac{17}{3} \, z^3 \right)
        \nonumber \\ &
        + z^2 \, \ln^2(z) \, (36+z^2) + (1-z^2) \, \left(\frac{17}{3} - \frac{64}{3} \, z +
        \frac{17}{3} \, z^2 \right) \, \ln (1-z)
        \nonumber \\ &
        - 4 \, (1+30 \, z^2 + z^4) \, \ln(z) \ln(1-z) -(1+16 \, z^2 +z^4)  \left( 6 \, \mbox{Li}(z) - \pi^2 \right)
        \nonumber \\ &
        - 32 \, z^{3/2} (1+z) \left[\pi^2 - 4 \, \mbox{Li}(\sqrt{z})+ 4 \, \mbox{Li}(-\sqrt{z})
        - 2 \ln(z) \, \ln \left( \frac{1-\sqrt{z}}{1+\sqrt{z}} \right) \right] \, .
\end{align}
We stress that the function $f(z)$ refers to on-shell renormalization of the charm quark mass.

\begin{figure}[t]
\begin{center}
    \includegraphics[width=8cm]{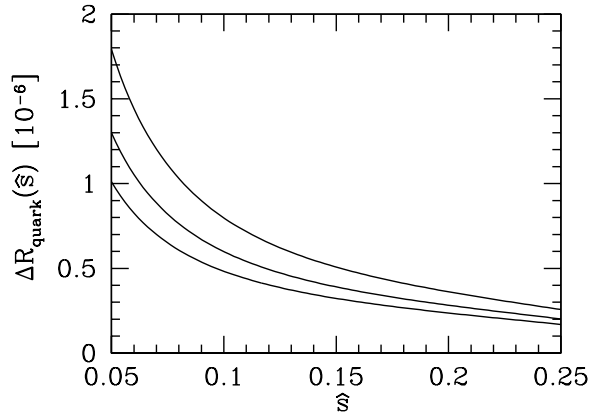}
    \caption{The new contribution $\Delta R_{\text{quark}}(\s)$ due to
     finite bremsstrahlung corrections for $\mu=2.5$ GeV (uppermost curve),
     $\mu=5$ GeV (middle curve) and $\mu=10$ GeV (lowest curve) and $m_c/m_b=0.29$.}
    \label{fig:deltar}
\end{center}
\end{figure}

\begin{figure}[t]
\begin{center}
    \includegraphics[width=8cm]{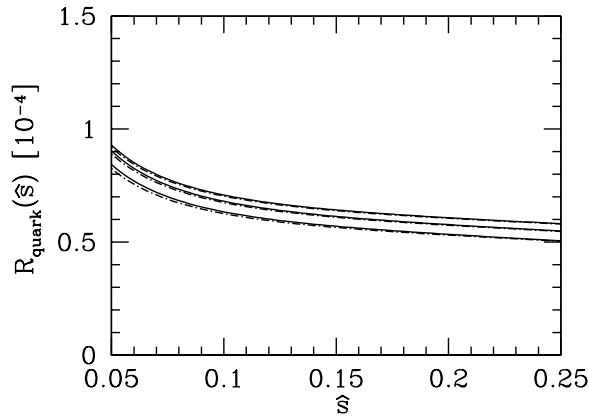}
    \caption{The solid curves show the ratio $R_{\text{quark}}(\s)$
     including the finite bremsstrahlung corrections while the
     dash-dotted curves show the corresponding results without the new
     corrections. The uppermost curves (solid and dash-dotted) correspond to
     $\mu=10$ GeV, the middle curves to $\mu=5$ GeV and the lowest
     curves to $\mu=2.5$ GeV. $m_c/m_b=0.29$.
    \label{fig:ratio}}
\end{center}
\end{figure}

In Fig.~\ref{fig:deltar} we consider the contribution $\Delta R_{\text{quark}}(\s)$, which is due to the finite
bremsstrahlung corrections in Eqs.~(\ref{finitea}) and (\ref{eq:tau}), for three values of the renormalization scale
($\mu$=2.5, 5 and 10 GeV) and for fixed value $m_c/m_b=0.29$. The values of all the other input parameters are as in
Ref.~\cite{virtCorr}.
In Fig.~\ref{fig:ratio} we combine the new corrections with the previous results. The solid lines show the ratio
$R_{\text{quark}}(\s)$, including the new corrections, for the values $\mu=10$~GeV (uppermost curve), $\mu=5$ GeV
(middle curve) and $\mu=2.5$ GeV (lowest curve) and for fixed value $m_c/m_b=0.29$. The dashed lines represent the
corresponding results without the new corrections. We find that for $\s=0.05$ the new corrections increase the ratio
$R_{\text{quark}}(\s)$ by $\sim 3\%$, while for larger values of $\s$ their impact is even smaller.
When including the finite bremsstrahlung corrections we obtain
\begin{equation*}
\label{rint}
    R_{\text{quark}} = \int\limits_{0.05}^{0.25} \!d\s \,
    R_{\text{quark}}(\s) = (1.27 \pm 0.08 (\mu)) \times 10^{-5}
\end{equation*}
{}for the integrated quantity $R_{\text{quark}}$. The error is obtained by varying $\mu$ between 2.5 GeV and 10 GeV.
For comparison, the corresponding result without the finite bremsstrahlung correction is $R_{\text{quark}}(\s) = (1.25
\pm 0.08 (\mu)) \times 10^{-5}$ \cite{virtCorr}.

Among the errors on  $R_{\text{quark}}(\s)$ due to uncertainties in the input parameters, the one related to the charm
quark mass is by far the largest. We therefore only comment on this error. In principle, the uncertainties induced by
the charm quark mass have two sources. First, it is unclear whether $m_c$ in the virtual- and bremsstrahlung
corrections should be interpreted as the pole mass or the $\overline{\mbox{MS}}$ mass (at an appropriate scale).
Second, the question arises what the numerical value of $m_c$ is, once a choice concerning the definition of $m_c$ has
been made.

To illustrate these problems more clearly, it is useful to first consider the process $B \to X_s \gamma$. There, the
one-loop matrix elements of $O_1$ and $O_2$ vanish, implying that the charm quark mass dependence only enters at
$\Order(\alpha_s)$. Formally, one can interpret $m_c$ in these $\Order(\alpha_s)$ expressions to be the pole mass or
the $\overline{\mbox{MS}}$ mass because the difference is of higher order in $\alpha_s$. Nevertheless, it has been
argued in the literature \cite{Gambino01} that the choice $m_c^{\overline{\rm{MS}}}(\mu)$ with $\mu \in [m_c,m_b]$
seems more reasonable than $m_c^{\rm{pole}}$ (which was used in all the previous analysis) due to the fact that the
largest charm quark mass dependence comes from the real part of the two-loop matrix elements of $O_1$ and $O_2$, where
the charm quarks are usually off-shell, with a momentum scale set by $m_b^{\rm{pole}}$ (or some seizable fraction of
it). It was shown in Ref.~\cite{Gambino01} that the definition of the charm quark mass leads to a relatively large
uncertainty in the branching ratio: Changing $m_c/m_b$ in $\Gamma(B \to X_s \gamma)$ from $0.29 \pm 0.02$ to $0.22 \pm
0.04$, i.e., from $m_c^{\rm{pole}}/m_b^{\rm{pole}}$ to $m_c^{\overline{\rm{MS}}}/m_b^{\rm{pole}}$ (with $\mu \in
[m_c,m_b]$), causes an enhancement of $\mbox{BR}(B \to X_s \gamma)$ by $\sim 11\%$.

In the process $B \to X_s\, \ell^+ \ell^-$ this problem is less severe because $m_c$ enters already the one-loop
diagrams (i.e., at $\Order(\alpha_s^0)$) associated with $O_1$ and $O_2$. As the two-loop calculation requires the
renormalization of $m_c$, the definition of $m_c$ has to be specified. Therefore, the two-loop result explicitly
depends on the definition of the charm quark mass. This can be seen from \cite{virtCorr}. For the pole mass definition,
the results for the two-loop matrix elements of $O_1$ and $O_2$, encoded in $F_{1,2}^{(7),(9)}$, are given in
Eqs.~(54)--(56), while those corresponding to the $\overline{\mbox{MS}}$ definition are obtained by adding the terms
$\Delta F_{1,2,m_c \rm{ren}}^{\rm{ct} (9)}$ given in Eq.~(49).

In the following, we investigate the impact of pole- vs. $\overline{\mbox{MS}}$ definition of $m_c$ in the rare decay
$b \to X_s\, \ell^+ \ell^-$ on the ratio $R_{\text{quark}}(\s)$. In the semileptonic decay $b \to X_c\, e\, \bar \nu_e$
the charm quark is basically on-shell. Therefore, we always use the pole mass definition for the charm quark mass in
$\Gamma(b \to X_c\, e \,\bar \nu_e)$, which enters  $R_{\text{quark}}(\s)$.

\begin{figure}[t]
\begin{center}
    \includegraphics[width=8cm]{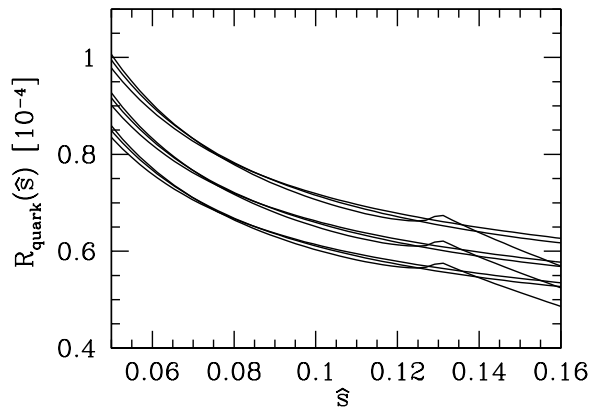}
    \caption{$R_{\text{quark}}(\s)$ for various values and definitions
    of $m_c$: The three bands are obtained by
    setting $m_c^{\rm{pole}}/m_b^{\rm{pole}}$=0.31 (uppermost), 0.29 (middle)
    and 0.27 (lowest) in $\Gamma(b \to X_c\, e\, \bar \nu_e)$. In the rare decay
    $b \to X_s\, \ell^+ \ell^-$ we set $m_c^{\overline{\rm{MS}}}/m_b^{\rm{pole}}=0.18,0.22,0.26$. This leads to three
    curves all within a narrow band. See text.}
    \label{fig:massdef1}
\end{center}
\end{figure}
In Fig.~\ref{fig:massdef1} we set $m_c^{\rm{pole}}/m_b^{\rm{pole}}$ equal to 0.31, 0.29 and 0.27 in the decay width
$\Gamma(b \to X_c\, e\, \bar \nu_e)$. In the rare decay $b \to X_s\, \ell^+ \ell^-$, on the other hand, we use the
$\overline{\mbox{MS}}$ definition for $m_c$, and put $m_c^{\overline{\rm{MS}}}/m_b^{\rm{pole}}=0.18,0.22$ and 0.26
(independently of $m_c^{\rm{pole}}/m_b^{\rm{pole}}$, to be on the conservative side). This leads, for a given value of
$m_c^{\rm{pole}}/m_b^{\rm{pole}}$, to three curves which form a narrow band. The uppermost band corresponds to
$m_c^{\rm{pole}}/m_b^{\rm{pole}}=$ 0.31, the middle to 0.29 and the lowest to 0.27. The curves with the strange
behavior for $\s>0.13$ all belong to the lowest value $m_c^{\overline{\rm{MS}}}/m_b^{\rm{pole}}=0.18$. As the result
for the two-loop corrections was derived in expanded form which only holds for $\s < 4\, m_c^2/m_b^2$, the strange
behavior illustrates that, for $m_c/m_b=0.18$, the result is not valid for $\s > 0.13$.

\begin{figure}[t]
\begin{center}
    \includegraphics[width=8cm]{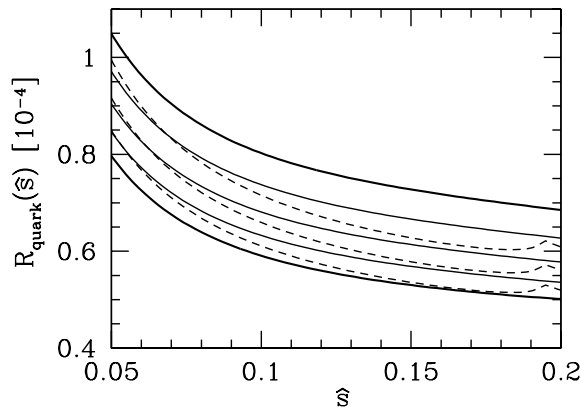}
    \caption{$R_{\text{quark}}(\s)$ for various values and definitions
    of $m_c$: The solid curves are obtained by
    setting $m_c^{\rm{pole}}/m_b^{\rm{pole}}$=0.33 (uppermost), 0.31, 0.29, 0.27
    and 0.25 (lowest) in the rare- and the semileptonic decay.
    The dashed lines are obtained by taking
    $m_c^{\overline{\rm{MS}}}/m_b^{\rm{pole}}=$ 0.22 in the rare decay
    and $m_c^{\rm{pole}}/m_b^{\rm{pole}}$= 0.31, 0.29 and 0.27 in
    $\Gamma(b \to X_c\, e\, \bar \nu_e)$. See text.
    \label{fig:massdef2}}
\end{center}
\end{figure}
In Fig.~\ref{fig:massdef2} the three middle solid curves are obtained by adopting the pole mass definition of $m_c$,
both in the rare and in the semileptonic decay. They correspond to $m_c^{\rm{pole}}/m_b^{\rm{pole}}=$ 0.31, 0.29, 0.27.
The dashed curves, on the other hand, are obtained when the $\overline{\mbox{MS}}$ definition with
$m_c^{\overline{\rm{MS}}}/m_b^{\rm{pole}}=$ 0.22 is used in the rare decay width. One finds that for $\s>0.06$ the
results for $R_{\text{quark}}(\s)$ are somewhat larger when using the pole mass definition of $m_c$ in the rare decay.
For values below $\s<0.06$ the situation is reversed and thus the same as for $b \to X_s \gamma$ \cite{Gambino01}.
Again, the strange behavior of the dashed curves indicates that, for $m_c/m_b=0.22$, the expanded formulas become
unreliable for values of $\s > 0.19$ . The thick solid lines are obtained by adopting the pole mass definition on the
whole and correspond to $m_c/m_b$=0.33 (upper) and 0.25 (lower). In summary, the figure shows that the quark mass
uncertainties can effectively be estimated by working with the pole mass definition throughout, provided one takes the
rather conservative range $0.25 \le m_c^{\rm{pole}}/m_b^{\rm{pole}} \le 0.33$.

\begin{figure}[t]
\begin{center}
    \includegraphics[width=8cm]{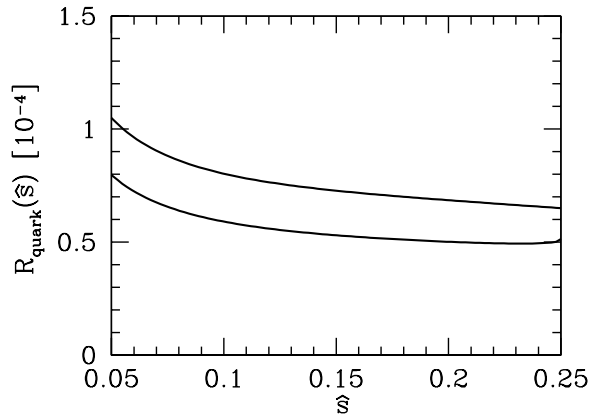}
    \caption{$R_{\text{quark}}(\s)$ for
    $m_c^{\rm{pole}}/m_b^{\rm{pole}}$=0.33 (uppermost), 0.31, 0.29, 0.27 and 0.25 (lowest) in the rare- and the
    semileptonic decay in the full window $\s \in [0.05,0.25]$.
    \label{fig:massdef3}}
\end{center}
\end{figure}
Finally, in Fig.~\ref{fig:massdef3} we show $R_{\text{quark}}(\s)$ in the full range $\s \in [0.05,0.25]$ for
$m_c^{\rm{pole}}/m_b^{\rm{pole}} \in [0.25,0.33]$. Note that for these values of $m_c/m_b$ the expanded formulas hold
just up to $\s=0.25$.

Comparing Fig.~\ref{fig:ratio} with Fig.~\ref{fig:massdef3}, we find that the uncertainty due to $m_c/m_b$ is clearly
larger than the leftover $\mu$ dependence. Varying $m_c/m_b$ between 0.25 and 0.33, the corresponding uncertainty
amounts to $\pm 15 \%$.

To conclude: We have calculated the finite gluon bremsstrahlung corrections of $\Order(\alpha_s)$ to $\Gamma(\bsll)$,
taking into account the contributions of the operators $O_1$, $O_2$, $O_7$, $O_8$, $O_9$ and $O_{10}$. We have worked
out the numerical impact of the new corrections on the invariant mass spectrum of the lepton pair in the range $\s \in
[0.05,0.25]$ and found an increase of about 3\% for $\s=0.05$ and even less for larger values of $\s$. Furthermore, we
investigated the uncertainties of $R_{\text{quark}}(\s)$ due to the definition and numerical uncertainties of the charm
quark mass. We found that these errors can be reliably estimated when working with the pole mass definition of $m_c$,
provided one takes the rather conservative range $0.25 \le m_c^{\rm{pole}}/m_b^{\rm{pole}} \le 0.33$.

Acknowledgements: \\
The work of H.M.A. was partially supported by NATO Grant PST.CLG.978154.

%
%
\newpage
\appendix
\section{Auxiliary quantities $A_i$, $T_9$, $U_9$ and $W_9$}
The auxiliary quantities $A_i$, $T_9$, $U_9$ and $W_9$, appearing in the modified Wilson coefficients in Eq.
(\ref{modcoeff}) and in the effective Wilson coefficients in Eqs.~(\ref{effcoeff7})--(\ref{effcoeff10}) are the
following linear combinations of the Wilson coefficients $C_i(\mu)$ \cite{Bobeth:2000,ALGH}:
    \begin{align}
        \label{ATUW}
        A_7 =&\, \frac{4\, \pi}{\alpha_s(\mu)} \, C_7(\mu) - \frac{1}{3} \, C_3(\mu) - \frac{4}{9} \, C_4(\mu) -
                \frac{20}{3} \, C_5(\mu) - \frac{80}{9} \, C_6(\mu) \, ,  \nonumber \\
        A_8 =&\, \frac{4\, \pi}{\alpha_s(\mu)} \, C_8(\mu) +  C_3(\mu) - \frac{1}{6} \, C_4(\mu) + 20 \, C_5(\mu) -
                \frac{10}{3} \, C_6(\mu) \, ,  \nonumber \\
        A_9 =&\, \frac{4 \pi}{\alpha_s(\mu)} \, C_9(\mu) + \sum_{i=1}^{6} \, C_i(\mu) \, \gamma_{i9}^{(0)} \,
                \ln \! \left( \frac{m_b}{\mu} \right)
                 + \frac{4}{3} \, C_3(\mu) + \frac{64}{9} \, C_5(\mu) + \frac{64}{27} \, C_6(\mu) \, ,  \nonumber \\
        A_{10} =&\, \frac{4 \pi}{\alpha_s(\mu)} \, C_{10}(\mu) \, , \\
        T_9 =&\, \frac{4}{3} \, C_1(\mu) +  C_2(\mu) + 6 \, C_3(\mu) + 60 \, C_5(\mu) \, ,
                \nonumber \\
        U_9 =& - \frac{7}{2} \, C_3(\mu) - \frac{2}{3} \,C_4(\mu) -38 \,C_5(\mu) - \frac{32}{3} \,C_6(\mu) \, , \nonumber \\
        W_9 =& - \frac{1}{2} \, C_3(\mu) - \frac{2}{3} \,C_4(\mu) -8 \,C_5(\mu) - \frac{32}{3} \,C_6(\mu) \, .\nonumber
    \end{align}
The elements $\gamma_{i9}^{(0)}$ can be found in \cite{Bobeth:2000}, while the loop-function $h(z,\s)$ is given in Eq.
(\ref{hfun}).

In the contributions which explicitly involve virtual or bremsstrahlung correction only the leading order coefficients
$A_i^{(0)}$, $T_9^{(0)}$, $U_9^{(0)}$ and $W_9^{(0)}$ enter. They are given by
    \begin{align}
        \label{ATUW0}
        A_7^{(0)} =&\, \, C_7^{(1)} - \frac{1}{3} \, C_3^{(0)} - \frac{4}{9} \, C_4^{(0)} -
                \frac{20}{3} \, C_5^{(0)} - \frac{80}{9} \, C_6^{(0)} \, ,  \nonumber \\
        A_8^{(0)} =&\, \, C_8^{(1)} +  C_3^{(0)} - \frac{1}{6} \, C_4^{(0)} + 20 \, C_5^{(0)} -
                \frac{10}{3} \, C_6^{(0)} \, ,  \nonumber \\
        A_9^{(0)} =&\, \frac{4\, \pi}{\alpha_s} \left( C_9^{(0)} + \frac{\alpha_s}{4\,\pi}\, C_9^{(1)} \right) +
                \sum_{i=1}^{6} \, C_i^{(0)} \, \gamma_{i9}^{(0)} \,
                \ln \! \left( \frac{m_b}{\mu} \right) + \frac{4}{3} \, C_3^{(0)} + \frac{64}{9} \, C_5^{(0)} +
                \frac{64}{27} \, C_6^{(0)} \, ,  \nonumber \\
        A_{10}^{(0)} =&\, C_{10}^{(1)} \, , \vspace*{0.3cm} \\ \vspace*{0.3cm}
        T_9^{(0)} =&\, \frac{4}{3} \, C_1^{(0)} +  C_2^{(0)} + 6 \, C_3^{(0)} + 60 \, C_5^{(0)} \, ,
                \nonumber \\
        U_9^{(0)} =& - \frac{7}{2} \, C_3^{(0)} - \frac{2}{3} \,C_4^{(0)} -38 \,C_5^{(0)}
                - \frac{32}{3} \,C_6^{(0)} \, , \nonumber \\
        W_9^{(0)} =& - \frac{1}{2} \, C_3^{(0)} - \frac{2}{3} \,C_4^{(0)} -8 \,C_5^{(0)}
                - \frac{32}{3} \,C_6^{(0)} \, .\nonumber
    \end{align}
We list the leading and next-to-leading order contributions to the quantities $A_i$, $T_9$, $U_9$ and $W_9$ in Tab.
\ref{table3}.
\renewcommand{\st}{\rule[-2.5mm]{0mm}{8mm}}
\def\lb{\raisebox{0.5mm}{\big(}}
\def\rb{\raisebox{0.5mm}{\big)}}
\begin{table}[hbt]
    \begin{center}
    \begin{tabular*}{\textwidth}{l@{\hspace*{1.5cm}}c@{\hspace*{1.5cm}}c@{\hspace*{1.5cm}}c}
        \hline\hline
        \st$\mu$                                & $ 2.5$ GeV          & $ 5$ GeV           & $ 10$ GeV           \\
        \hline
        \st$\alpha_s                          $ & $ 0.267           $ & $ 0.215          $ & $ 0.180           $ \\
        \st$C_1^{(0)}                         $ & $ -0.697          $ & $ -0.487         $ & $-0.326           $ \\
        \st$C_2^{(0)}                         $ & $ 1.046           $ & $ 1.024          $ & $ 1.011           $ \\
        \st$\lb A_7^{(0)},~A_7^{(1)}\rb       $ & $ (-0.360,~0.031) $ & $(-0.321,~0.019) $ & $ (-0.287,~0.008) $ \\
        \st$A_8^{(0)}                         $ & $ -0.164          $ & $ -0.148         $ & $ -0.134          $ \\
        \st$\lb A_9^{(0)},~A_9^{(1)}\rb       $ & $ (4.241,~-0.170) $ & $(4.129,~0.013)  $ & $ (4.131,~0.155)  $ \\
        \st$\lb T_9^{(0)},~T_9^{(1)}\rb       $ & $ (0.115,~0.278)  $ & $(0.374,~0.251)  $ & $ (0.576,~0.231)  $ \\
        \st$\lb U_9^{(0)},~U_9^{(1)}\rb       $ & $ (0.045,~0.023)  $ & $(0.032,~0.016)  $ & $ (0.022,~0.011)  $ \\
        \st$\lb W_{9}^{(0)},~W_{9}^{(1)}\rb   $ & $ (0.044,~0.016)  $ & $(0.032,~0.012)  $ & $ (0.022,~0.009)  $ \\
        \st$\lb A_{10}^{(0)},~A_{10}^{(1)}\rb $ & $ (-4.372,~0.135) $ & $(-4.372,~0.135) $ & $ (-4.372,~0.135) $ \\
        \hline\hline
    \end{tabular*}
    \end{center}
\caption{\label{table3} Coefficients appearing Eqs.~(\ref{effcoeff7})--(\ref{effcoeff10}) for $\mu=2.5$~GeV,
$\mu=5$~GeV and $\mu=10$~GeV. For $\alpha_s(\mu)$ (in the $\overline{\mbox{MS}}$ scheme) we used the two-loop
expression with five flavors and $\alpha_s(m_Z)=0.119$. The entries correspond to the pole top quark mass
$m_t=174$~GeV. The superscript (0) refers to lowest order quantities.}
\end{table}
%
%

\end{document}